\documentstyle[12pt]{article}
 \catcode`\@=11

\newskip\humongous \humongous=0pt plus 1000pt minus 1000pt

\newif\ifdtup

\def\ie{\hbox{\it i.e.}{}}	\def\etc{\hbox{\it etc.}{}}
	
\def\etal{\hbox{\it et al.}}    

\def\beq{\begin{equation}}
\def\eeq{\end{equation}}

\def\beqn{\begin{eqnarray}}
\def\eeqn{\end{eqnarray}}
\relax
\let\beqa=\beqn
\let\eeqa=\eeqn

\relax

\def\dotx{\dotx{\dot\overline{x}}}

\def\prd#1#2#3{
	{\it Phys. Rev. }{\bf D~#1} (19#3) #2}

\def\pra#1#2#3{
	{\it Phys. Rev. }{\bf A~#1} (19#3) #2}

\relax

\catcode`\@=11

\def\@normalsize{\@setsize\normalsize{15pt}\xiipt\@xiipt
\abovedisplayskip 14pt plus3pt minus3pt%
\belowdisplayskip \abovedisplayskip
\abovedisplayshortskip  \z@ plus3pt%
\belowdisplayshortskip  7pt plus3.5pt minus0pt}

\def\small{\@setsize\small{13.6pt}\xipt\@xipt
\abovedisplayskip 13pt plus3pt minus3pt%
\belowdisplayskip \abovedisplayskip
\abovedisplayshortskip  \z@ plus3pt%
\belowdisplayshortskip  7pt plus3.5pt minus0pt

\def\@listi{\parsep 4.5pt plus 2pt minus 1pt
            \itemsep \parsep
            \topsep 9pt plus 3pt minus 3pt}}

\def\underline#1{\relax\ifmmode\@@underline#1\else
	$\@@underline{\hbox{#1}}$\relax\fi}
\@twosidetrue

\catcode`\@=11

\catcode`\@=12

\relax

\def\figcap{\section*{Figure Captions\markboth
	{FIGURECAPTIONS}{FIGURECAPTIONS}}\list
	{Fig. \arabic{enumi}:\hfill}{\settowidth\labelwidth{Fig.
999:}
	\leftmargin\labelwidth
	\advance\leftmargin\labelsep\usecounter{enumi}}}
 \relax
\def\tablecap{\section*{Table Captions\markboth
	{TABLECAPTIONS}{TABLECAPTIONS}}\list
	{Table \arabic{enumi}:\hfill}{\settowidth\labelwidth{Table
999:}
	\leftmargin\labelwidth
	\advance\leftmargin\labelsep\usecounter{enumi}}}
 \relax
\def\reflist{\section*{References\markboth
	{REFLIST}{REFLIST}}\list
	{[\arabic{enumi}]\hfill}{\settowidth\labelwidth{[999]}
	\leftmargin\labelwidth
	\advance\leftmargin\labelsep\usecounter{enumi}}}
 \relax

\catcode`\@=11

\def\IFPUB{}
\def\IFPub#1{\def\IFPUB{#1}}
\def\ps@headings{
\def\@oddhead{\hbox{\rm M. Moreno \it et al.}\hfill
	\makebox[.5\textwidth]{\raggedright\ignorespaces
--\thepage{}--
	\hfill {\rm IFUNAM -- \IFPUB}  }
	     }
\def\@evenhead{\@oddhead}
\def\@oddfoot{}
\def\@evenfoot{\@oddfoot}
\def\subsectionmark##1{\markboth{##1}{}}
}

\ps@headings

\IFPub{FT94-45}
\relax

\newbox\negrabox
\def\negra#1{{\setbox\negrabox=\hbox{$#1$}
   \kern-.025em\copy\negrabox\kern-\wd\negrabox
   \kern.05em\copy\negrabox\kern-\wd\negrabox
   \kern-.025em\raise.0433em\box\negrabox}}

\newcommand{\half}{\frac{1}{2}}

\newcommand{\bdm}{\begin{displaymath}
                 }
\newcommand{\edm}{\end{displaymath}
                 }

\newcommand{\kn}{{\bf k}}

\newcommand{\ku}{{\kn \uparrow}}

\newcommand{\kl}{{\kn \lambda}}
\newcommand{\kli}[1]{{\kn_{#1} \lambda_{#1}}}

\catcode`@=12

\relax

\relax

\begin{document}

\thispagestyle{empty}

\begin{flushright}
\small Preprint IFUNAM-\IFPUB \\
\tiny	version 2.1, Nov.  8 1993.
version 2.2, March 25,1994.
\end{flushright}
\vspace{-1.1cm}

\begin{center}
{\bf \large \bf
\noindent
  Exotic and conventional superconductivity in a Dirac supersymmetric
scheme }
 \end{center}
 \begin{quote}
 {
 M.~Moreno$^a$,
\\
 R.~M.~M\'endez-Moreno, S.~Orozco and  M.~A.~Ort\'{\i}z$^b$,
\\
    M.~de~Llano$^{c}$
 }
\\[0.12cm]
 {\it
 $^a$.Instituto de F\'{\i}sica,
Universidad Nacional Aut\'onoma de M\'exico, \\
Apartado Postal 20-364, 01000 M\'exico, D.~F., MEXICO
\\ $^b$. Deptartamento de F\'{\i}sica, Facultad de Ciencias,\\
Universidad Nacional Aut\'onoma de M\'exico, \\
Apartado Postal 21-092, 04021 M\'exico, D.~F., MEXICO
 \\ $^c$. Physics Department, North Dakota State University,
\\Fargo, ND 58105, USA
}
\\[0.25cm]
\end{quote}

\vspace*{-1.6cm}
\parindent 2.2em

\noindent
 {\bf Abstract.}\ A new pairing theory for many-fermion systems is
obtained via the Dirac supersymmetry framework recently introduced to
describe Dirac particles in external potentials.  It is shown that
the standard Bogoliubov-Valatin canonical transformation treatment of
the quasi-particle BCS singlet pairing mechanism naturally falls
within this framework.  Straightforward generalizations in which the
fermions can be ascribed  $ \nu$ {\it components} are shown to lead
to enhanced gap energies and critical temperatures as in the case of
cuprate superconductors without invoking a stronger electron-boson
coupling.  The new  $T_{c}^{max}$ limit is $T_c^{max} =   \nu
T_c^{BCS}$, with
$T_c^{BCS} \approx  40^0 K$.

\vfill

\pagebreak[3]
\vspace{0.5cm} \noindent
	One of the most striking approaches to the problem of pairing
was pioneered by Nambu~\cite{nambu:8589} who showed that a remarkable
connection exists between the pairing mechanism and
supersym\-metry\cite{wess:83}. Within this framework, supersymmetry
must be severely broken in order to make contact with phenomenology.
It has recently been shown, see
Refs.~\cite{moreno:8990,martinez:91,moreno:92}, that a generalization
of quantum supersymmetry to massive Dirac
fermions~\cite{kostelecky:85} can be achieved if the Dirac
Hamiltonian is of the form
 \beq
 \label{eq:Dsusy}
  H = Q + Q^\dagger + \Lambda,
 \eeq
 with $\Lambda$ a Hermitean operator, provided that these operators
satisfy the following anticommutation relations
 \beq
 \label{eq:Qlam}
 \{ Q , \Lambda \}= \{ Q^\dagger , \Lambda \} = 0 = Q^2.
 \eeq
 This new structure has been named {\it Dirac supersymmetry} (Dirac
Susy).

 	In this letter we apply Dirac Susy concepts to
superconductivity phenomena and demonstrate that one may accommodate
within this formalism both the usual BCS mechanism as well as other
kinds of fermion pairing.  In view of the generality of the approach,
it is likely that the underlying mechanisms of the so-called exotic
superconductors\cite{uemura:91} can be treated within the scope of
Dirac Susy. One indication of this is seen in the example of a
two-layered fermion gas in which a substantial increase in the energy
gap can be achieved even for weak interaction. Our formalism lends
itself to study the consequences of a BCS mechanism for a general
fermion system with several components within the canonical
transformation scheme proposed by Bogoliubov and Valatin
(BV)~\cite{BV:58}.  For a recent survey of the relevance of BCS-based
mechanisms to high-temperature superconductivity (HTSC) see
Ref.\cite{ginzburg:92}.  The notion of {\it component} that we use
should be understood in a broad sense: it can arise physically from a
spin degree of freedom, or from a pseudo-spin generated by layers in
which an electron or hole moves; it can also stem from a finite
number of transversal fermion states in a slab of finite thickness;
or in general from any condition leading to a {\it discrete}
representation of an degree of freedom. The latter is the case of the
electron gas~\cite{takada:92} in a doped multi-valley semiconductor
or
multi-valley semimetal.

	 A Dirac Susy structure such as (\ref{eq:Dsusy}) admits an
exact unitary Foldy-Wouthuysen (FW) transformation of the original
Hamiltonian~\cite{moreno:8990,martinez:91,moreno:92}. The squared
Hamiltonian becomes
$ H^2 = \{ Q , Q^\dagger \} + \Lambda^2 \equiv h^2 + \Lambda^2$,
  where $h$ is required to be an even root of the supersymmetric
operator, $ h^2 = \{ Q , Q^\dagger \}$ in the FW sense. Defining
 $\hat{\Lambda} \equiv {\Lambda}/{\sqrt{\Lambda^2}},$ the FW
transformed Hamiltonian is then
 \beq  \label{eq:hfw}
 H_{FW} = \hat{\Lambda} \sqrt { \{ Q , Q^\dagger \} + \Lambda^2 },
\eeq
 provided $\Lambda$ has non-null eigenvalues. Further, $Q$ and
$Q^\dagger$ in matrix form are
 \beq \label{eq:A1}
 \begin{array}{cc}
 Q = \left(
 \begin{array}{cc}
 0 & 0 \\
 q & 0
 \end{array}
 \right)
, & \mbox{~~}
Q^\dagger = \left(
 \begin{array}{cc}
 0 & q^\dagger \\
 0 & 0
 \end{array}
 \right)
 \end{array},
\eeq
 where the $q$ are in general submatrices. Because $\hat{\Lambda}$
commutes with $H_{FW}$ and its eigenvalues are $\pm 1,$
Eq.~(\ref{eq:hfw}) implies generally that, the energy spectra has two
branches; between them there is a gap determined by the lowest
eigenvalues of $\Lambda^2$ and $h^2$.  If $\hat{\Lambda}$ has
negative eigenvalues and $H^2$ is not bounded from above one gets an
unstable ground state for the Dirac Susy Hamiltonian
(\ref{eq:Dsusy}). The usual remedy is to introduce a Dirac, or Fermi,
sea.  The important feature of Dirac Susy interactions is that the
definition of this sea is coupling strength independent, this
property being the {\it stability of the Dirac
sea}~\cite{martinez:91}. The introduction of the sea necessarily
implies a field theory.

     Note that a standard quantum mechanical Susy is approximately
obtained if $\Lambda^2$ eigenvalues are either very large or very
small as compared to those of $h^2$. Therefore, Dirac Susy is a
specific type of supersymmetry-breaking.
 It is more appropriate to consider it as a generalization of the
usual supersymmetry to cases when the Hamiltonian is fermion-like.  A
relevant consequence to our purpose here is that Dirac Susy restores
and guarantees a gap in the energy spectrum of the system.

	Let us first consider how this formalism is connected to the
BV transformation in BCS pairing phenomena.  The essential
characteristic of BV theory is to exhibit the well-known Fermi sea
{\it instability}\cite{schrieffer:64} through a canonical
transformation which mixes fermion (electron, for definiteness) and
hole states of opposite momentum and spin quantum numbers ( viz.,
${\kn \uparrow}$ and $-{\kn \downarrow}$). The physical basis for
this transformation relies on the fact that individual electrons in
the Fermi sea near the Fermi surface escape the sea and become bound
into Cooper pairs. The new states, so-called {\it bogolons}, are
related to the electron states, $ a_\ku$ and $\ a_{-{\kn
\downarrow}}, $ by a transformation matrix which must be orthogonal
in order to preserve the anticommutation relations for the bogolon
operators, $\alpha_\kn$ and $\beta_{-{\kn}}$.  We follow
the notation of Fetter and Walecka~\cite{fetter:71}.
 The new vacuum state $| {\bf 0}\!\!>$ satisfies
 $
 \alpha_\kn| {\bf 0}\!\!> \, = \, \beta_\kn| {\bf 0}\!\!> \,= \, 0.
 $
A realization of this new vacuum could be, {\it e.g.}, the BCS state
  $
 | {\bf 0}\!\!> \, = \, {\cal N}\ \prod_{i}{( 1 + g_i a_i^\dagger
a_{-i}^\dagger}) | \phi >,
 $
  where $ | \phi > $ is the original (quasifermion) vacuum, $ {\cal
N}$
is a normalization factor and $g_i$ is related to the amplitude of
pairing in the $i$ and $-i$ states. Here, $i$ stands for the relevant
single-particle quantum numbers. Clearly, the new vacuum, $| {\bf
0}\!\!>,$ is not annihilated by the quasifermion operators $a_i$.
 At zero temperature the thermodynamic potential~\cite{fetter:71}
$\Omega(T=0,V,\mu)$ is given by the expectation value of the operator
${K} = {H} - \mu {N}$
 \beqa
 {K} & = &
      \sum_{{\kn \lambda}} ( \epsilon^0_k - \mu) a_\kl^{\dagger}
a_\kl \\
 \nonumber
 & & - \half  \mbox{\hspace{-1.5em}}\sum_{%
 \mbox{\footnotesize $
			 \begin{array}{c}
			{\bf \footnotesize k}_1 , {\bf \footnotesize
k}_2  ,
			{\bf \footnotesize k}_3 , {\bf \footnotesize
k}_4
			\\
			 \lambda_1, \lambda_2,  \lambda_3, \lambda_4
	       				\end{array}
  $}}
\mbox{\hspace{-3.0em}}
< \kn_1 \lambda_1 \kn_2 \lambda_2 | V | \kn_3 \lambda_3 \kn_4
  \lambda_4 >
\mbox{\hspace{0.2em}}a_\kli{1}^{\dagger} a_\kli{2}^{\dagger}
a_\kli{4} a_\kli{3}.
  \eeqa
 Applying the BV transformation to this leads to separation of the
operator ${K}$ into a zero-body, $H_0$, a one-body, $H_I$, and a
two-body, $H_{II},$ operators; the weak-coupling assumption allows
treating $H_{II}$ as a small perturbation to be neglected.
$ H_I   $ written in a Nambu\cite{nambu:8589} matrix form is

 \beq  \label{eq:nambu}
 H_I = \sum_\kn \left(
 \begin{array}{c}
 \alpha_\kn \\
 \beta_{-\kn}^\dagger
 \end{array}
 \right)^\dagger \left(
 \begin{array}{cc}
 u_k & v_k \\
 - v_k & u_k
 \end{array}
 \right) \left(
 \begin{array}{cc}
 \xi_k & \Delta_k \\
 \Delta_k & - \xi_k
 \end{array}
 \right) \left(
 \begin{array}{cc}
 u_k & -v_k \\
 v_k & u_k
 \end{array}
 \right) \left(
 \begin{array}{c}
 \alpha_\kn \\
 \beta_{-\kn}^\dagger
 \end{array}
 \right).
  \eeq
 where $\xi_k$ measures the Hartree-Fock quasi-particle energy
$\epsilon^0_k$ with respect to the chemical potential $\mu$ and
$\Delta_k$ is the gap function
 $
 \Delta_k = \sum_{k'} < k -k | V | k' -k'> u_k v_k.
 $ As will be shown shortly, (\ref{eq:nambu}) already implies a Dirac
Susy structure.

  The BV transformation is fixed by the requirement that this
one-particle Hamiltonian $H_I$ should be diagonal for a nonzero value
of $ v_k$ with respect to the $\alpha, \beta$ basis. The bogolon
excitation energy is then given by
 $
	E_k = \sqrt{\xi_k^2 +  \Delta_k^2},
	$
\  and the gap function
$\Delta_k$ is determined from the self-consistent equation
 \mbox{
 $
 \Delta_k
       = \half \sum_{k'} < k -k | V | k' -k'> {\Delta_{k'}}/{E_{k'}}.
 $
 }
 \  The superconducting solution being characterized by $\Delta_k
\neq 0$.
The familiar BCS interaction model giving a simple yet realistic
nontrivial solution is of the form
 $
 < k -k | V | l -l >\  = ({V_0}/{L^d}) \theta (\hbar \omega_D -
|\xi_k|)
 \theta (\hbar \omega_D - |\xi_l|),
 $
  \ where $V_0$ is the (square of the) electron phonon coupling,
$L^d$ in $d$ dimensions, is a normalization volume and $\omega_D$ is
the cutoff (Debye) frequency . Introducing $N(0)$, the density of
states for one spin projection at the Fermi surface, one obtains the
celebrated {\it gap equation,}
 \beq
 1  = \frac{ V_0 N(0)}{2} \int_{- \hbar \omega_D}^{\hbar \omega_D}
\frac{d \xi}{\sqrt{\Delta^2 +
  \xi^2}}
   =  V_0 N(0)\sinh^{-1}(\hbar \omega_D/\Delta),
 \eeq
 Letting $V_0 N(0) = g$ and
 solving for $\Delta$ one gets
 $
 \Delta =  \hbar \omega_D / \sinh (1 / g)
 \approx 2 \hbar \omega_D
 e^{-1/g},
 $
 for weak coupling $g \ll 1$. For pure elements $g$ is in the
interval $ 0.15 \leq g \leq 0.6 $.  In a finite-temperature formalism
this implies
 $
 T_c \approx 1.13 \hbar \omega_D e^{-1/g}.
 $
 \

 	We now generalize the BV transformation for a system in which
each fermion is characterized by a discrete parameter $\lambda$ that
can take on $\nu$ values.  For $\nu =2$, this could correspond to the
two spin-$\frac {1}{2}$ degrees of freedom. Alternatively, if the
electrons are confined to a slab of thickness $t$, the discreteness
in $\lambda$ corresponds to the different excitations of the
transverse degree of freedom and in addition to the spin. Other
concrete examples are a many layered electron gas, for which
$\lambda$ refers to the specific layer, and the multi-valley
structures referred~\cite{takada:92} to before.
	In close analogy with the one-component case let us now
define
 \beq
 A_k^T \equiv (
 {a_k}_1,  \ \cdots
 \ {a_k}_\nu, \ {a_{-k}}_1^\dagger , \ \cdots
 \ {a_{-k}}_\nu^\dagger
),
 \eeq
where $T$ stands for transpose matrix, and
 \beq \label{eq:U}
 B_k \equiv \left(
 \begin{array}{c}
 \check{\alpha}_{{\bf k}} \\ \check{\beta}_{-{\bf k}}^\dagger
 \end{array}
 \right) = \left(
 \begin{array}{cc}
  \check{u}_k & \check{v}_k \\ -\check{v}_k & \check{u}_k^*
 \end{array}
 \right) A_k = U A_k,
 \eeq
 where $\nu \times \nu$ matrices and $\nu$-vectors are denoted with a
check above. The commutation relations for the operators
${a_k}_\lambda$
can be reexpressed in matrix form as
 \beq
 \label{eq:canon}
 A_k A_{k'}^\dagger \pm A_{k'}^* A_k^T = 1 \delta_{k k'},
 \eeq
 while the canonicity condition of the BV transformation requires
that
 $B_k B_{k'}^\dagger \pm B_{k'}^* B_k^T = 1 \delta_{k k'}$,
 which is satisfied if $U$ in (\ref{eq:U}) is an orthogonal matrix,
as can easily be verified by sandwiching (\ref{eq:canon}) between $U$
and $U^\dagger$. When this transformation is applied to the
Hamiltonian $K$, one obtains as before an operator of the form $K =
H_0 + H_I +H_{II}$. Again the term $H_{II}$ is neglected. One has for
$H_I$ the form (\ref{eq:nambu}) but with all quantities bearing a
check, \ \ie, they are $\nu \times \nu$ matrices.
  For our purposes the relevant structure is just
 \beq
 \left(
 \begin{array}{cc}
 \check{\xi}_k & \check{\Delta}_k \\ \check{\Delta}_k & -
\check{\xi}_k
 \end{array}
 \right).
 \eeq
  If in (\ref{eq:Dsusy}) we identify
 \beq
 Q = \left(
 \begin{array}{cc}
 0 & 0 \\
 \check{q} & 0
 \end{array}
 \right) = \left(
 \begin{array}{cc}
 0 & 0 \\
 \check{\Delta}_k & 0
 \end{array}
 \right),
\mbox{~~and ~~}
 \Lambda = \left(
 \begin{array}{cc}
 \check{\xi}_k & 0 \\
 0 & - \check{\xi}_k
 \end{array}
 \right),
 \eeq
 one can easily confirm our central result that the standard BV
operator and its extrapolations to multicomponent systems have in
fact a Dirac Susy structure, and therefore it guarantees an energy
gap. In this case, the BV transformation is completely equivalent to
the FW transformation. Quite generally the condition (\ref{eq:Qlam})
implies
 \beq
  \label{eq:Dexi}
  [\check\Delta,\check\xi] = 0,
 \eeq
 which is satisfied if \  $\check \xi = \xi_0 \check 1 + \xi_1 \check
\Delta$ , where $\xi_0$ and $ \xi_1$ are functions of $k$.

  Let us now introduce a finite-temperature formalism, it is useful
to consider a mean-field approximation~\cite{reichl:80,sigrist:91}.
  The thermal expectation values are
$
 < {\cal A} > =   Tr (e^{- \beta K} {\cal A})/Z,
$ with $Z \equiv Tr (e^{- \beta K} ),$
 one has in this context that
 $
 < a_i^\dagger a_{-i}^\dagger> \ \neq 0 \neq \ < a_i a_{-i}>
 $
 where we have merged the indices $k$ and $\lambda$ into the single
$i$. With this notation, the matrix elements of $\check \Delta$ in
the finite temperature formalism are
 \beq
 \Delta_{\lambda_1 \lambda_2} (k) = \sum_{l \lambda_3 \lambda_4}{
  <\lambda_1 k ; \lambda_2 -k | V | \lambda_3 l; \lambda_ 4 -l>
< a_{-l \lambda_4} a_{l \lambda_3} >
}.
\eeq
An effective one-body  Hamilton operator can be written as
 \begin{equation}
 \tilde K = \sum_{i}{ [ \epsilon (k) - \mu ] a_i^\dagger a_{i}}
   + \half \sum_{i j}^{}{ [\Delta_{i j} (k) a_i^\dagger
a_{-i}^\dagger
+h.c.]} \ ,
 \end{equation}
 which leads to the Dirac Susy structure in the finite temperature
formalism
 \begin{equation}
 \label{eq:rrr}
  \tilde K = \sum_{k}^{}{ A_k^\dagger
 \left [\matrix{
   [\epsilon (k) - \mu ] \check 1& \check \Delta(k) \cr
 \check \Delta^\dagger(k) &  - [ \epsilon (k) - \mu ] \check 1 \cr
 }\right ]
 A_k},
\end{equation}
 as in the $H_I$ case.  Then we have found Dirac Susy structure in
the finite temperature formalism in a system with multicomponentes,
independent of the structure and the intensity of the coupling.

 	  We now illustrate the general ideas developed above with
the
following examples with $\nu = 2$ for which the components are not
(physically) equivalent to spin. This in practice means that the spin
degree of freedom will be assumed to couple to a singlet state in the
Cooper pair.  The physical systems to which these models correspond
are in fact inspired in the quasi-2D electron gas. a) First, consider
an electron gas confined to a 2D slab of finite thickness; we assume
that only the two lowest transversal energy states are important,
\ie, the energy separation between the successive transversal states
is big enough so that higher states can be neglected. b) Secondly,
consider a two-layer 2D electron gas for which the spin degree of
freedom could be neglected, and take the case for which the fermion
degrees of freedom have been diagonalized {\it before} the BV
transformation. These therefore describe quasiparticles {\it \`a la }
Landau, and as emphasized by Schrieffer~\cite{schrieffer:64} the
pairing mechanism is more adequately applied to these states.
Formally these states are described by diagonal matrices $ \check
\xi_k $. In both these models we take $ \check \xi_k = \check 1
\xi_k; $ physically this means that the two components of (quasi-)
electrons are equivalent before the BV transformation. Consequently
(\ref{eq:Dexi}) is satisfied for an arbitrary matrix $\check\Delta_k
$. Performing the BV transformation on (\ref{eq:rrr}) we {\it derive}
the bogolon Hamiltonian
\beq
 H_{BV} =
   \pmatrix{
\sqrt{ \xi_k^2 + \check{\Delta}_k \check{\Delta}_k ^\dagger} & 0 \cr
   0
 & - \sqrt{ \xi_k^2  + \check{\Delta}_k ^\dagger  \check{\Delta}_k}
\cr
           } \ .
 \eeq
 This form is further diagonalized by any transformation
diagonalizing $\check{\Delta}_k \check{\Delta}_k ^\dagger$ and $
\check{\Delta}_k ^\dagger \check{\Delta}_k$, it being sufficient to
study either one of these two matrices because they appear in
diagonal blocks and have the same eigenvalues. It is convenient to
expand $\check\Delta$ in terms of Pauli matrices,
${\check\negra{\tau}},$ namely:
 $
  \check{\Delta} = \Delta_0 \ {\check 1} + {\bf {\Delta}} \cdot
  {\check\negra{\tau}},
 $
 \ with ${\bf {\Delta}}$ a 3-vector, where the subindex $k$ has been
omitted, and $\Delta_0$ and ${\bf {\Delta}}$ might be complex. Then,
 \begin{eqnarray}
  \check{\Delta}\check{\Delta} ^\dagger
        &=&    [ \Delta_0  \Delta_0^*
                   + {\bf{\Delta}} \cdot {\bf {\Delta}^*}]  \ {\check
1}
 \nonumber  \\
          & &    +     [   ( \Delta_0 {\bf {\Delta}^*}  +
	  	\Delta_0^* {\bf {\Delta}} )
                   + i {\bf {\Delta}} \times {\bf {\Delta}^*}]
		   \cdot {\check\negra{\tau}},
 \end{eqnarray}
 where the term proportional to ${\check\negra{\tau}}$ is also real.
Furthermore, the two vectors
 $( \Delta_0 {\bf {\Delta}^*} + \Delta_0^*
{\bf {\Delta}} )$
 and
 $ {\bf {\Delta}} \times {\bf {\Delta}^*}$
 are clearly orthogonal to each other. Consequently the diagonal form
of
$  \check{\Delta}\check{\Delta} ^\dagger$,
$  (\check{\Delta}\check{\Delta} ^\dagger)_d$,
 is simply
 \begin{equation} \label{eq:Ad}
   (\check{\Delta}\check{\Delta} ^\dagger)_d
  =
    [ \Delta_0 \Delta_0^* + {\bf {\Delta}} \cdot
    	{\bf {\Delta}^*}] {\check 1 }
      +  \sqrt{ ( \Delta_0  {\bf {\Delta}^*}  +  \Delta_0^*
      	 {\bf {\Delta}} )^2
      +   |{\bf {\Delta}} \times
      	  {\bf {\Delta}}^*|^2} \  {\bf {\check\tau}}_3.
 \end{equation}
 Now, the lowest (highest) transition temperature $T_c$ is
essentially determined by the smallest (largest) eigenvalue of $
\check{\Delta}_k$, via $\Delta(T_c)=0$.  Eq.(\ref{eq:Ad}) is the most
general form of the gap equation for the effective Hamiltonian of
Eq.(\ref{eq:rrr}), with $\nu = 2$.

  Let us now discuss some concrete realizations of this formalism.
First, it follows that the conventional BCS result can be obtained
with
${\bf {\Delta}}=0$ and $\Delta_0 = \delta$. This gap form is
consistent with
$ V_{\lambda_1 \lambda_2 \lambda_3 \lambda_4,k l } =
\delta_{\lambda_1 \lambda_2} \delta_{\lambda_3 \lambda_4}
V_{kl} $.
For this case a finite temperature formalism gives the usual phase
diagram and the ratio
$R=  2 \Delta(T=0)/ (k_b T_c) = 3.1$,

A different {\it Ansatz} which is important in connection
with the exotic heavy fermion phenomena~\cite{sigrist:91} is found
assuming
$\Delta_0 = 0$ and real ${\bf{\Delta}}$ (up to an overall phase).
 Such a situation can be obtained dynamically with an
electron-electron interaction of the form
 $ V_{\lambda_1 \lambda_2 \lambda_3 \lambda_4, k l }
 = 1/2 (\delta_{\lambda_1 \lambda_3} \delta_{\lambda_2 \lambda_4}
 - \delta_{\lambda_1 \lambda_4} \delta_{\lambda_2 \lambda_3})
 V_{kl} $.
 If all the components are of
the same size $\delta$ (which physically means that all fermion
couplings leading to Cooper pairing are equivalent) we get an {\it
enhancement} of the gap for this triplet interaction,
 $\Delta(0) = \sqrt{3} \delta$.
  In this case one gets a phase space diagram which is of the same
generic form of the traditional BCS but for which
 $\Delta(0) =  \sqrt{3} \Delta(0)_{BCS} $;
 that implies  an increased
 $T_c =  \sqrt{3} T_{c BCS} $.
 A new physical situation arises for the case of $\Delta_0 =
\Delta_i$ but for this the enhancement is greater, $T_c = 2 T_{c
BCS}$.  This means that triplet and singlet Cooper pairs are combined
and that they add constructively to $T_c$.
 With a greater number of components, $\nu$, and assuming a equal
coupling among them, we conjecture that even larger gap enhancements
of order $\nu$ can be obtained if a material in which all the
components interact with similar couplings.

 A different effect arises if one seeks a big splitting in
(\ref{eq:Ad}). This is exemplified with $\Delta_0 =0 $ and {\it one}
cartesian component of ${\bf{\Delta}}$ being imaginary relative to
the
other two, for example $\Delta_1 = \Delta_2 = i \Delta_3 = \delta$,
with $\delta$ real.  Under such conditions there are two
well-differentiated values for the gap $ (3 \pm \sqrt{8}) \delta^2;$
which in turn imply two transition temperatures.  In a model with
more components one expects a multi-gap system with multi-transition
temperatures.  In this discussion we have mainly been concerned with
discrete fermion degrees of freedom. We remark, however, that this
generally has implications for the orbital pair wave functions. For
example, in the layered realization mentioned above, coupling among
fermions in different layers will require a spherically {\it
asymmetric} wave function, which in turn will imply P, D,
\etc \ wave components~\cite{goss:93}.

 In conclusion, the  general Dirac Susy is adequate for pairing
phenomena. There exists a general connection between
Dirac supersymmetry and the BCS theory of superconductivity. Dirac
supersymmetry was sketched, in particular the remarkable result that
a unitary FW transformation decoupling positive and negative energy
states can be explicitly constructed.  The relation to conventional
non-Dirac supersymmetry, with a possible central charge extension,
was elucidated.  It is implicitly shown that a much more complex
supersymmetry breaking term is needed to understand superconductivity
than previously thought. However, this is naturally taken into
account by Dirac supersymmetry. So for the case of a singlet
conventional pairing theory we showed that the effective
quasi-electron Hamiltonian is precisely of the Dirac supersymmetric
form, and that the BV transformation lends itself naturally to its
derivation.
 The relation between superconductivity and Dirac supersymmetry was
then generalized to a multicomponent fermionic system. For a {\it
two}-component system it was shown that a simple {\it Ansatz} leads
to an increase in $T_c$ by a factor of $2$, and for a $\nu$ component
system the gap increase can conjectured to be of order
$\nu$.


\vspace{0.5cm}


This work is partially supported by Direcci\'on General de Asuntos
del Personal Acad\'emico, Project IN102991, Universidad Nacional
Aut\'onoma de M\'exico (U.N.A.M.), M\'exico, D.~F. , M\'exico. M. de
Ll. thanks NATO (Belgium) for a research grant.




\vspace{0.5cm}


\begin{thebibliography}{99}
\bibitem{nambu:8589} Y. Nambu, {\it Physica }{\bf D 95,} 147 (1985);
   M. Mukerjee and Y. Nambu, {\it Ann. Phys.} (N.Y.)  {\bf 191,}143
(1989).


\bibitem{wess:83} J.~Wess and J.~Bager, {\it Supersymmetry and
Supergravity}
		(Princeton Univ. Press, Princeton, N.~J., 1983).


\bibitem{moreno:8990}  M. Moreno and A. Zentella, {\em J. Phys.} {\bf
A 22}
			L821 (1989);
  M. Moreno, R. Mart\'{\i}nez and A. Zentella, {\em Mod. Phys. Lett.}
{\bf A
5,} 949 (1990).


\bibitem{martinez:91}  R. Mart\'{\i}nez, M.~Moreno and A. Zentella,
			\prd{43}{2036}{91}.

\bibitem{moreno:92}  M. Moreno and R.~M.~M\'endez-Moreno,  in
			{\it Proc. of the Workshop on High
			Energy Phenomenology,} Ed. by M.~A.~P\'erez
and
			R.~Huerta (World Scientific, Singapore, 1992)
p.~365;
			{\it ibid. } in
			{\it Symmetries in Physics,} Ed. by
			A.~Frank, and B.~Wolf, (Springer-Verlag,
1992)
			p.~185.


\bibitem{kostelecky:85} V.~A.~Kostelecky and M.~M.~Nieto
\pra{38}{4413}{88} and
references therein. These authors first showed that
 a massless Dirac electron in an external magnetic field exhibits
many
properties of supersymmetric quantum mechanics.


\bibitem{uemura:91} Y. E. Uemura \etal,
			{\it Phys. Rev. Lett.} {\bf 66,} 2665 (1991).


\bibitem{BV:58}  N.N. Bogoliubov, {\it \ Sov. Phys.-JETP } {\bf 7,}
			41 (1958);
   J.G. Valatin, {\it Nuovo Cimento } {\bf 7,} 843
			(1958).


 \bibitem{ginzburg:92} V. L. Ginzburg, {\it Physica}
 		{\bf C 209,} 1 (1992).


\bibitem{takada:92}
			Y.~Takada, {\it J.~Phys.~Soc.~Japan }{\bf
61}, 238(1992).


\bibitem{schrieffer:64}  J.~R.~Schrieffer, {\it Theory of
Superconductivity}
			(W.~A.~Benjamin,  New York, 1964).



\bibitem{fetter:71}  A. L. Fetter and J. D. Walecka,
			{\it Quantum Theory of Many-Particle Systems}
			(McGraw-Hill, New-York,1971).



\bibitem{reichl:80} L.~E.~Reichl, {\it A Modern Course in Statistical
Physics}
(University of Texas Press, Austin, 1980).

\bibitem{sigrist:91} M.~Sigrist and K.~Ueda,
		{\it Rev.~Mod.~Phys.} {\bf 63,} 239 (1991).

\bibitem{goss:93} B. Goss Levi, {\it Phys. Today} {\bf 46,} 17 (May,
1993).

\end{thebibliography}
\end{document}